\documentstyle[aps,prl,floats,graphicx,epsfig]{revtex}
\begin{document}
\draft

\twocolumn[\hsize\textwidth\columnwidth\hsize\csname @twocolumnfalse\endcsname

\title{Defect formation in inhomogeneous 2-nd order phase transition:
theory and experiment.}
\author{G.E. Volovik  }

\address{  Low Temperature Laboratory, Helsinki
University of Technology, Box 2200, FIN-02015 HUT, Espoo, Finland\\
and\\ Landau Institute for Theoretical Physics, Moscow, Russia}

\date{\today} \maketitle
\

\begin{abstract}
The status quo in our understanding of defect formation during a rapid
transition into the broken symmetry state in condensed matter and cosmology is
discussed. An observation of vortex nucleation in neutron
absorption experiments in superfluid $^3$He-B is interpreted in terms of
defect formation during inhomogeneous cooling through $T_c$. Due to the
temperature gradient in the locally heated region the superfluid phase
transition occurs as a propagating front. The theoretical considerations of
vortex formation at the propagating front are based on work by
Kibble-Volovik, Kopnin-Thuneberg, and
Aranson-Kopnin-Vinokur (AKV).
\end{abstract}
\
]

\section{Introduction}

To produce a new vortex line in the vortex-free state of superfluid liquid is
not an easy job. If the container is devoid of the remnant vorticity, which can
be pinned by rough surface,  the vortices are created only when a treshold
$v_c$ for  the hydrodynamic instability of the superflow is reached
\cite{SingleVortexNucleation}. The thermal activation or quantum tunneling can
assist the nucleation only in the narrow vicinity of the instability treshold,
where
 the external perturbations, however, are more effective. In superfluid
$^3$He-B,
because of the large  size $r_c$ of the vortex core,  the region near the
treshold, where thermal activation or quantum tunneling can be important, is
particularly small,  $v_c-v_s \sim 10^{-6}v_c$. In a typical cylindrical
container with radius $R=2.5$ mm and height $L=7$ mm, rotating with angular
velocity $\Omega= 3$ rad/s, the vortex-free state stores a huge amount of
kinetic energy
$(1/2)\int dV~\rho_s(v_s-v_n)^2\sim 10$ GeV. This energy cannot be
released, since
the intrinsic half period of the decay of this metastable state is essentially
larger than the proton life time.

\begin{figure}[!!!t]
\begin{center}
\leavevmode
\epsfig{file=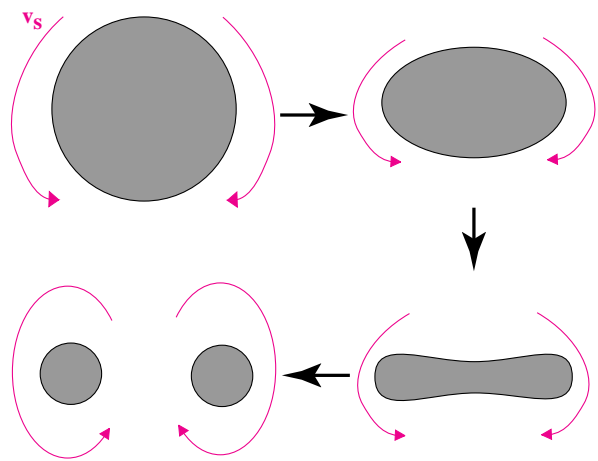,width=0.9\linewidth}
\caption{Evolution of the overheated normal fluid region into the core of the
vortex ring in the presence of the counterflow. View from 70's.}
\label{CounterflowScenario}
\end{center}
\end{figure}

That cosmic rays can assist in releasing this energy by producing vortex rings,
I first heard from my supervisor, professor Iordanskii, in 1972. The natural
scenario for that was thought as depicted in Fig.\ref{CounterflowScenario}. The
energetic particle heats a region above the superfluid transition temperature
$T_c$.  During the cooling the normal liquid in this region can continuously
evolve to form the core of a vortex loop, which starts growing if the radius
$R_b$ of the heated region is larger than the radius of the ring sustained by
counterflow, i.e. if
$|v_s-v_n| >  v_{c1}=   (\kappa/4\pi R_b)
\ln {(R_b/r_c)} $, where $\kappa$ is the quantum of circulation around the
vortex.

If the counterflow essentially exceeds this treshold, the evolution,
which is most favourable for vortex production, leads to the closely packed
vortex rings of the critical size, which can further develop. This gives the
following estimation for maximal number of vortex loops, which can grow
further:
$N\sim (|v_s-v_n| /  v_{c1})^3$.

Experiments with irradiated superfluid $^3$He were started in 1992
in Stanford, where it was found that the irradiation  assists the transition
of supercooled  $^3$He-A to $^3$He-B. In 1994  the neutron irradiation
of $^3$He-B was found to produce a shower of quasiparticles in Lancaster
\cite{Bradley} and vortices in  rotating $^3$He-B in Helsinki \cite{experiment}.
Energy deficit found in low-$T$ Grenoble experiments indicated possib
le
formation of vortices in $^3$He-B even without rotation \cite{Bunkov1}. In
Helsinki the  observed number of vortices produced per one event showed
both the
treshold behavior and the cubic dependence at large rotation velocity:
Above the
treshold it was well approximated by $N\sim (|v_s-v_n| /  v_{c1})^3 -1$.  This
indicated that the nature has chosen some scenario, which produces the maximal
possible number of vortices. What is the reason for that?

The decay products from the neutron absorption reaction generate
ionization tracks, the details of which are not
well known in liquid $^3$He. At the moment we have two working scenaria of
thermalization of the energetic particles:

(i) The mean free path is long and increases with decreasing of the energy.
This can lead to a ``Baked Alaska'' effect, as has been described by Leggett
\cite{Leggett2}. A thin shell of the radiated high energy particles expands
with the Fermi velocity $v_F$, leaving behind a region at reduced $T$. In this
region, which is isolated from the outside world
by a warmer shell, a new phase can be formed. Such  Baked
Alaska mechanism for generation of new phase has also been discussed in high
energy physics, where it describes the result of a hadron-hadron collision. In
this relativistic case the thin shell of energetic particles expands with the
speed of light. In the region separated from the exterior vacuum by the hot
shell a false vacuum with a chiral condensate can be formed
\cite{Bjorken}.   This scenario provides possible
explanation of formation of the B-phase in the supercooled A-phase
\cite{Leggett2}.

(ii) During thermalization the mean free path is less than the dimension of the
region where the energy is deposited and the temperature is well determined
during the phase transition through $T_c$.   In this case there is no
Baked-Alaska effect: no hot shell separating the interior region from
the exterior. So the exterior region can effectively fix the phase in
the cooled bubble, suppressing the formation of the vacuum states,
which would be different from that in the bulk liquid.  Due to this proximity
effect the formation of vortices can be also suppressed.

In both cases of monotonic and nonmonotonic temperature
profile, two mechanisms of the vortex formation are important:

(a) The Kibble-Zurek (KZ) mechanism of the defect formation during the
quench. For
the scenario (ii), where the interior region is not separated from
the exterior by the warmer shell, the KZ mechanism is to be modified
to include spatial inhomegeneity, which leads to the moving transition front.
The proximity effect of the exterior region is not effective if the
phase transition front moves sufficiently rapidly \cite{KV,KT,DLZ}. The
modified
KZ mechanism is not sensitive to the existence of the external counterflow,
which
only role is to extract the formed vortices from the bubble. The same
KZ mechanism could be responsible for the formation of the A-B interfaces,
which
provides another scenario of the B-phase nucleation in the supercooled A-phase
\cite{Volovik,Bunkov2}.

(b) Instability of the normal-superfluid interface, which occurs in the
presence
of the counterflow \cite{AKV}.

Here we discuss these two mechanisms (a) and (b) of  vortex formation during
inhomogeneous quench as manifested in numerical simulations \cite{AKV}.

\section{KZ scenario in pesence of planar front.}

For a rough understanding of the KZ scenario of vortex
formation let us consider the time-dependent Ginzburg-Landau (TDGL)
equation for the one-component order parameter (OP)
$\Psi=\Delta/\Delta_0$:
\begin{equation}
        \tau_0{\partial \Psi\over \partial t}= \left(1-{T({\bf r},t)\over
T_c}\right) \Psi - \Psi|\Psi|^2 + \xi_0^2 \nabla^2 \Psi ~~.
\label{TDGL}
\end{equation}
Here $\tau_0 \sim 1/\Delta_0$ and $ \xi_0$  are
correspondingly the relaxation time of the OP and the coherence
length far from $T_c$.

If the quench occurs homogeneously in the whole space
${\bf  r}$, the temperature depends only on one parameter, the quench time
$\tau_{\rm Q}$:
\begin{equation}
        T(t) \approx  \left(1-{t\over \tau_{\rm Q}}\right)T_c  ~~.
\label{Thomogeneous}
\end{equation}
In the presence of a temperature gradient, say, along $x$, a new
parameter appears:
\begin{equation}
        T(x-ut)\approx\left(1-{t-x/ u\over  \tau_{\rm Q}}\right) T_c
~~.
\label{TInPropFront}
\end{equation}
Here $ u$ is the velocity of the temperature front which is
related to the temperature gradient
\begin{equation}
        \nabla_x T={T_c \over  u  \tau_{\rm Q}}~~.
\label{TempGrad}
\end{equation}

There exists a characteristic critical velocity $
u_c$ of the propagating temperature front. At $u \geq u_c$ the vortices are
formed, while at $u \leq u_c$  the defect formation is either strongly
suppressed
\cite{KV,KT} or completely stops \cite{DLZ}.

At slow velocities, $u\rightarrow 0$, the order
parameter almost follows the transition temperature front:
\begin{equation}
      |\Psi(x,t)|^2 = \left(1-{T(x-ut)\over T_c}\right)~,~
T< T_c   ~~.
\label{PsiSlow}
\end{equation}
In this case the phase coherence is preserved behind the transition front and
thus no defect formation is possible.

\begin{figure}[!!!t]
\begin{center}
\leavevmode
\epsfig{file=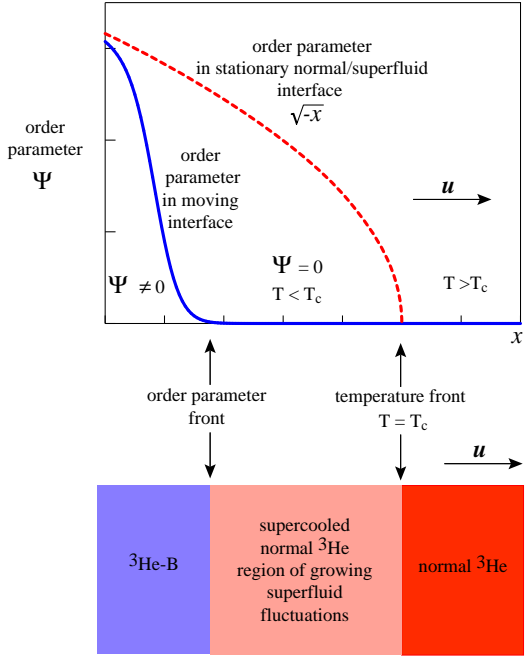,width=0.9\linewidth}
\caption{The OP distribution from Ref.\protect\cite{KV} at
nonzero velocity $u$  of the planar temperature front in the reference frame of
the moving front. The higher $u$ the larger is the lag between the temperature
front at
$x=0$, where $T=T_c$, and the OP front, which bounds the region with
$\Psi=0$. At
$u>u_c$ the exponentially growing OP fluctuations in the space
between these two boundaries are not washed out by the moving front and lead to
the vortex formation according to KZ scenario. Dashed line is the
Eq.(\protect\ref{PsiSlow}).}
\label{tdglfront}
\end{center}
\end{figure}

The extreme case of large velocity of the temperature front,
$u\rightarrow \infty$, corresponds to the homogeneous quench.
As was found by Kopnin and Thuneberg \cite{KT}, if $u$ is large
enough, the phase transition front cannot follow the temperature front: it lags
behind (see Fig.~\ref{tdglfront}). In the space between these two
boundaries the
temperature is already below the phase transition temperature, $T<T_c$, but the
phase transition did not yet happen, and the OP is still not formed,
$\Psi=0$. This situation is unstable towards the formation of bubbles of
the new
phase with $\Psi\neq 0$. This occurs independently in different
regions of the space, leading to vortex formation according to the KZ
mechanism. At a given point of space ${\bf r}$ the development of the
instability can be found from the linearized TDGL equation, since
during the intitial growth of the OP $\Psi$ the
cubic term can be neglected:
\begin{equation}
        \tau_0{\partial \Psi\over \partial t}=   { t \over
  \tau_{\rm Q}} \Psi  ~~.
\label{TDGLlinearized}
\end{equation}
This gives an exponentially growing OP, which starts from
some seed $\Psi_{\rm fluc}$, caused by fluctuations:
\begin{equation}
        \Psi({\bf r},t)=\Psi_{\rm fluc}({\bf r})\exp {t^2\over  2\tau_{\rm
Q}\tau_0 } ~~.
\label{ExponentialGrowth}
\end{equation}
Because of the exponential growth, even if the seed is small, the
modulus of the OP reaches its equilibrium value $|\Psi_{\rm eq}|
=\sqrt{1 -T/T_c}$ after the Zurek time $t_{\rm Z}$
\begin{equation}
       t_{\rm Z}=\sqrt{ \tau_{\rm
Q}\tau_0 }~~.
\label{ZurekTime}
\end{equation}
This occurs independently in different regions of space and thus the
phases of the OP in each bubble are not correlated. The
spatial correlation between the phases becomes important at distances
$\xi_{\rm v}$ where the gradient term in Eq. (\ref{TDGL}) becomes
comparable to the other terms at $t=t_{\rm Z}$. Equating the
gradient term $\xi_0^2 \nabla^2 \Psi \sim (\xi_0^2 /\xi_{\rm v}^2)
\Psi$ to, say, the term $\tau_0 \partial \Psi/\partial t|_{t_{\rm
    Zurek}} = \sqrt{\tau_0 / \tau_{\rm Q}}\Psi$, one obtains the
characteristic Zurek length scale which determines the initial
distance between the defects in homogeneous quench:
\begin{equation}
  \xi_{\rm v} =
  \xi_0 \; (\tau_{\rm Q}/\tau_0)^{1/4}.
\end{equation}

We can estimate the lower limit of the characteristic value of the fluctuations
$\Psi_{\rm fluc}=\Delta_{\rm fluc}/\Delta_0$, which serve as a seed for the
vortex formation. If there is no other source of fluctuations, caused, say, by
external noise, the initial seed is provided by thermal fluctuations of the
order
parameter in the volume $\xi_{\rm v}^3$.  The energy of such fluctuation is
$\xi_{\rm v}^3 \Delta_{\rm fluc}^2 N_F/E_F$, where $E_F$ is the Fermi
energy and
$N_F$ the fermionic density of states in the normal Fermi liquid. Equating this
energy to the temperature $T\approx T_c$ one obtains the magnitude of the
thermal
fluctuations of the OP
\begin{equation}
  {|\Psi_{\rm fluc}|\over |\Psi_{\rm eq}|} \sim  \left({\tau_0\over\tau_{\rm
Q}}\right)^{1/8} {T_c\over E_F}~.
\label{Fluctuations}
\end{equation}
Since the  fluctuations are initially rather small their growth time
exceeds the
Zurek time by the factor  $\sqrt{\ln  |\Psi_{\rm
eq}|/|\Psi_{\rm fluc}|}$.

  The criterium for the defect formation is
that the time of growth of fluctuations, $\sim ~t_{\rm Z}=\sqrt{ \tau_{\rm
Q}\tau_0 }$, is shorter than the time  $t_{\rm sw}=x_0(u)/u$ in which the
transition front sweeps the space between the two boundaries. Here
$x_0(u)$ is the lag between the transition temperature front and
the OP front (see Fig.~\ref{tdglfront}).
Thus the equation
$t_{\rm Z}= x_0(u_c)/u_c$
gives an estimate for the critical value $u_c$ of the velocity of
the temperature front, at which the laminar propagation becomes
unstable.  At large $u$ one has $x_0(u) \sim
u^3\tau_Q\tau_0^2/4\xi_0^2$ \cite{KT} and thus
\begin{equation}
   u_c\sim
{\xi_0\over \tau_0} \left({\tau_0\over\tau_{\rm Q}}\right)^{1/4}~,
\label{FrontCriticalVelocity}
\end{equation}
which agrees with estimation  $u_c=\xi_{\rm
v}/t_{\rm Z}$ in \cite{KV}.

 In the case of the neutron bubble the velocity of the temperature
front is
$u
\sim R_{\rm
  b}/\tau_{\rm Q}$, which makes $u \sim 10$ m/s. The critical
velocity $u_{\rm c}$ we can estimate to possess the same order
of magnitude value. This estimation suggests that the thermal gradient
should be sufficiently steep in the neutron bubble such that defect
formation can be expected. The further fate of the vortex tangle
formed under the KZ mechanism is the phase ordering process:
the intervortex distance continuously increases until it reaches the critical
size, when the vortex loops are expanded by the counterflow. This reproduces
the most favourable scenario of the vortex formation with the cubic law.

\section{Instability of normal/superfluid interface.}

\begin{figure}[!!!t]
\begin{center}
\leavevmode
\epsfig{file=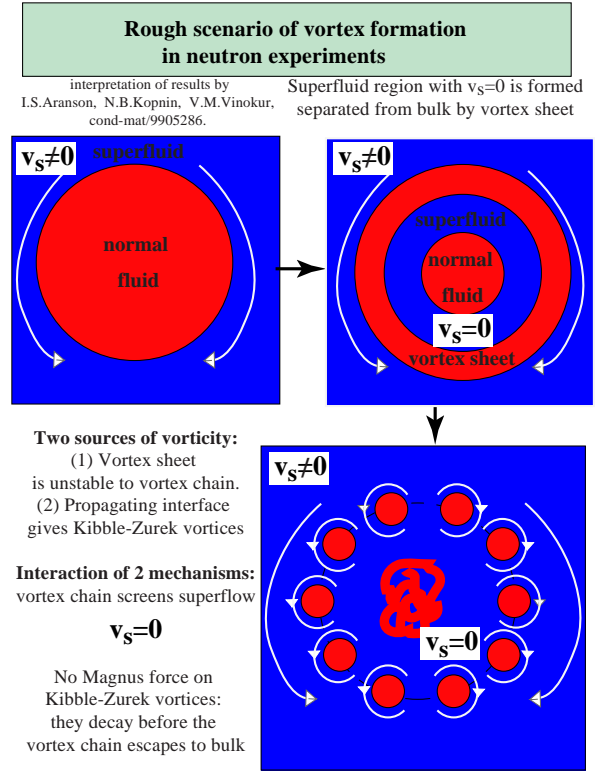,width=0.9\linewidth}
\caption{Rough scenario of instability of the superfluid-normal interface in
the presence of external superflow.}
\label{VortSheetInstability}
\end{center}
\end{figure}

Another mechanism of the vortex formation  has been recently found in 3D
numerical simulations in Ref.\cite{AKV}. It is related to the instability
of the
normal-superfluid interface  in the presence of the superflow. Let us
consider a
simple hand-waving interpretation of such an instability. The process can be
roughly splitted into two stages (see Fig.~\ref{VortSheetInstability}).

At first  stage the heated region of the normal liquid surrounded by the
superflow undergoes a superfluid transition. The transition should occur into
the state with the lowest energy, which corresponds to the  superfluid at rest,
i.e. with ${\bf v}_s=0$. Thus there appears the superfluid-superfluid
interface,
which separates the state with superflow (outside) from the state without
superflow (inside). Such a superfluid-superfluid interface with tangential
discontinuity of the superfluid velocity represents a vortex sheet by
definition.  Such vortex sheet, on which the phase of the OP is not
determined, was suggested by Landau and Lifshitz for He-II to describe the
superfluid state  of $^4$He under rotation \cite{LandauLifshitz} (see also
\cite{Onsager} and \cite{London}).

\begin{figure}[!!!t]
\begin{center}
\leavevmode
\epsfig{file=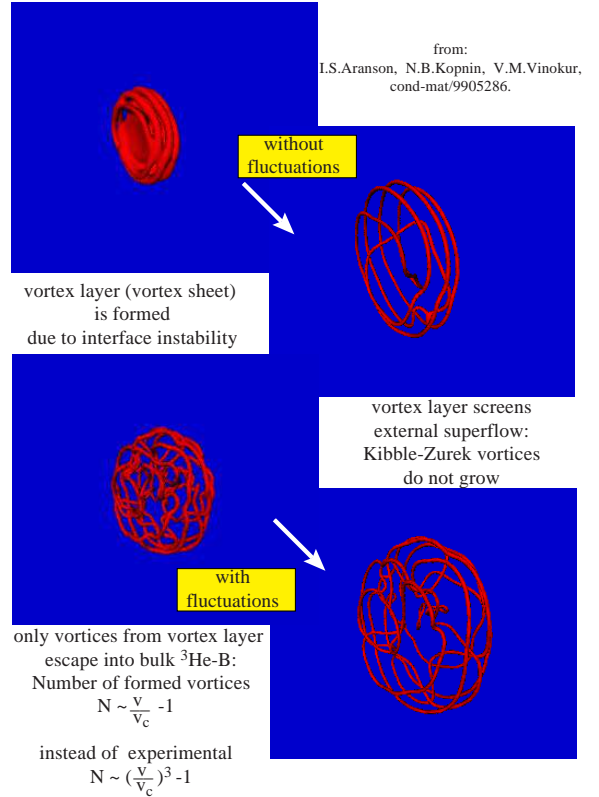,width=0.9\linewidth}
\caption{ 3D numerical TDGL simulations of vortex formation after the heated
region is cooled down (from Ref.~\protect\cite{AKV}).}
\label{NumSim1}
\end{center}
\end{figure}

\begin{figure}[!!!t]
\begin{center}
\leavevmode
\epsfig{file=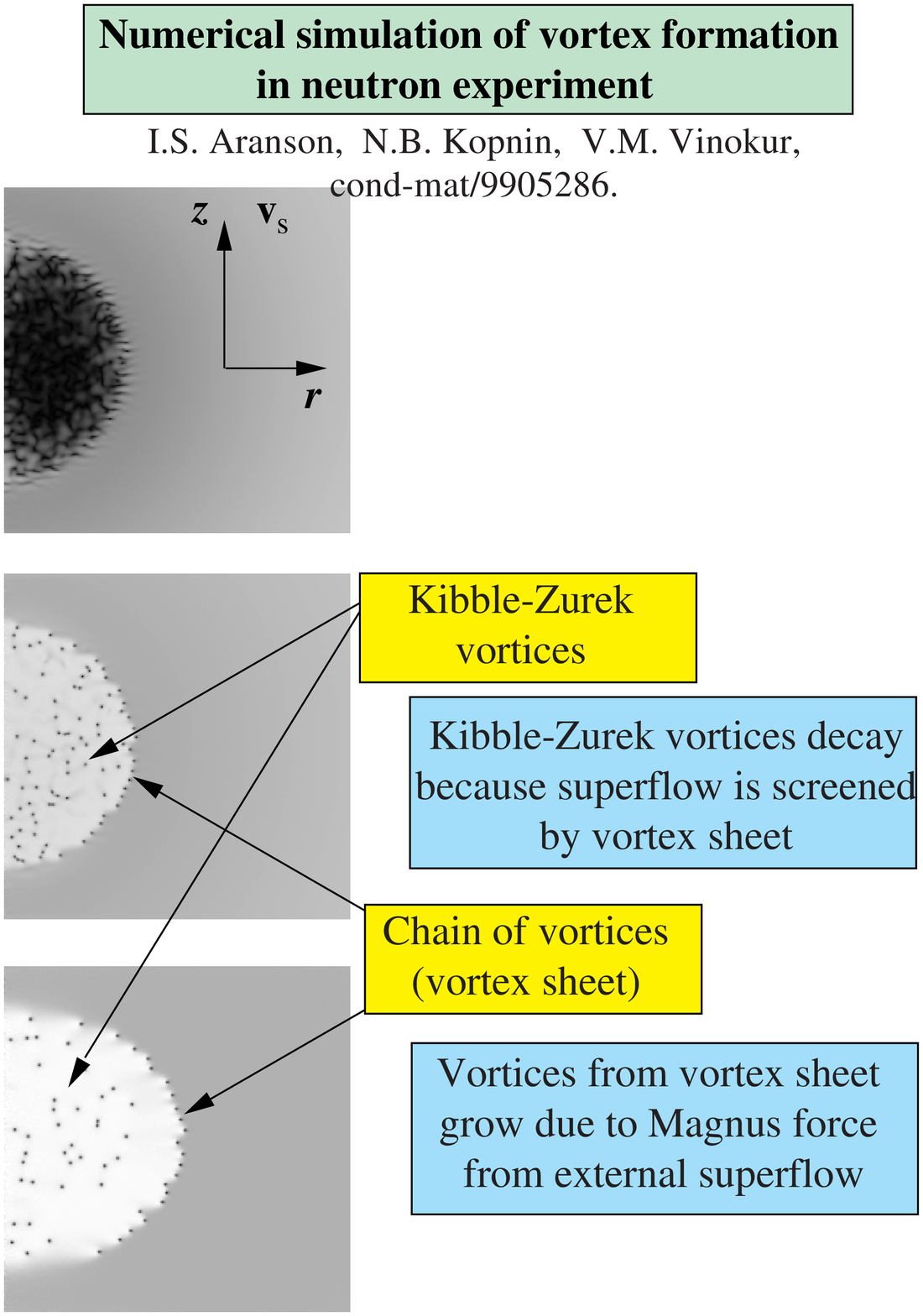,width=0.9\linewidth}
\caption{2D simulations, which  assume an axisymmetric development
(from Ref.~\protect\cite{AKV}).}
\label{NumSim2}
\end{center}
\end{figure}

The vortex sheet is unstable towards breaking up into a
chain of quantized vortex lines. The development of this instability
represents the second stage of the process. In numerical  simulation the
resulting chain of vortices is clearly seen (see
Figs.~~\ref{NumSim1} and \ref{NumSim2}).


The evolution in Fig.~\ref{CounterflowScenario} is thus caused by the
hydrodynamic instability of the normal/superfluid interface in the presence of
the tangential flow. Since vorticity is quantized, such instability leads
to the
formation of the vortex chain only above the treshold required to achieve the
circulation quantum from the tangential superflow.  If the  counterflow is
large
the number of vortices in this chain   $N\approx  |v_s -v_n| R_b /\kappa$, i.e.
one has a linear law instead of cubic.

Nucleation of the  KZ vortices due the motion of the
superfluid/normal interface is also observed in numerical simulations in
Ref.~\cite{AKV}. It occurs during shrinking of the interior region with
normal fluid.

\section{Discussion.}

Two mechanisms of the vortex formation have been identified  in
numerical simulations \cite{AKV}:  (a) vortices are formed behind the
propagating front due to KZ mechanism, as discussed in
Refs.\cite{KV,KT,DLZ}; and
in addition (b)  vortices are formed due to the corrugation instability (vortex
sheet  instability) of the front in the presence of external superflow.
Each of these mechanisms can be derived either analytically  for a simple
geometry, or understood qualitatively  with simple physical picture in mind.
The AKV calculations actually showed that each mechanism is fundamental: it
does
not depend much on the geometry and on parameters  of the TDGL equation.
Probably
both mechanisms hold even if TDGL theory cannot be applied.

The interplay of the  two mechanisms must depend on
details of the microscopic physics. In  their calculations
based on TDGL model, AKV found that the chain of vortices formed in the process
(b) screens the external superflow very effectively. The KZ vortices formed
in the
process (a) cannot grow: they decay before the screening chain
escapes to the bulk liquid. Thus in the AKV scenario only the chain of vortices
survives. This gives the linear dependence of the vortex number $N$ on the
counterflow $v_s - v_n$ instead of the observed cubic law.

This does not exclude the possibility  of another regime, where
KZ vortices have enough time to escape to the bulk. This is probably what the
cubic law found in Helsinki experiments tells us. Maybe the latter regime
cannot be obtained in the TDGL scheme and one must discuss the combined
dynamics
of the OP and quasiparticles.

In conclusion, in the period between  LT-21 and LT-22 the principles of
defect formation in inhomogeneous phase transition have been developed.


I am indepbt to N. Kopnin and M. Krusius for discussions
and I. Aranson and E. Thuneberg who kindly provided me with Figures from
their papers.


\end{document}